%% file: main_acm.tex
\documentclass[format=sigconf, review=false, anonymous=false, authorversion=true]{acmart}

\copyrightyear{2026}
\acmYear{2026}
\setcopyright{cc}
\setcctype{by-nc-nd}
\acmConference[UIST '26]{The 39th Annual ACM Symposium on User Interface Software and Technology}{November 02--05, 2026}{Detroit, MI, USA}
\acmBooktitle{The 39th Annual ACM Symposium on User Interface Software and Technology (UIST '26), November 02--05, 2026, Detroit, MI, USA}
\acmDOI{10.1145/3830398.3830509}
\acmISBN{979-8-4007-2856-3/2026/11}

\usepackage{svg}
\usepackage{booktabs} 
\usepackage{float}
\usepackage{listings}
\restylefloat{table}

\input{_setup.tex} 

\begin{document}

\tolerance=400 

%
\title[IntentLint]{IntentLint: Supporting Intent Scaffolding and Prompt-time Linting in Human-AI Collaborative Data Analysis}



\author{Felicia Li Feng}
\orcid{0000-0002-6198-0896}

\affiliation{%
  \institution{School of Computer Science, University of Waterloo}
  \country{}
}
\email{f4feng@uwaterloo.ca}

\author{Jian Zhao}
\orcid{0000-0001-5008-4319}

\affiliation{%
  \institution{School of Computer Science, University of Waterloo}
  \country{}
}
\email{jianzhao@uwaterloo.ca}

\author{Anamaria Crisan}
\orcid{0000-0003-3445-3414}

\affiliation{%
  \institution{School of Computer Science, University of Waterloo}
  \country{}
}
\email{ana.crisan@uwaterloo.ca}

\renewcommand{\shortauthors}{Feng et al.}

\begin{abstract}
In human-AI collaborative data analysis, as analyses rapidly evolve, the artifacts meant to capture shared understanding often become incomplete or difficult to interpret, leading to undocumented assumptions, cross-user misaligned intent, context-poor prompts, and unwanted agent behaviors. 
To address these challenges, we introduce \rv{a rule-based coordination layer with two interaction mechanisms,} intent scaffolding and prompt-time linting, that make analytic intent explicit and actionable during human–AI collaborative data analysis. 
We implement them in \sys{}, a proof-of-concept system that infers analytic intent from shared notebooks, represents it as structured, editable rules, and checks users' prompts against shared rules. \sys{} helps analysts externalize and refine their intent and proactively checks prompts for potential conflicts. 
A study with 16 data analysts shows that IntentLint improves awareness of collaborators’ intent and encourages reflection on analytic strategies, and provides design implications for supporting more aligned and transparent human-AI collaborative data analysis.
\end{abstract}

%
%



\begin{teaserfigure}
\centering
    \includegraphics[width=1\linewidth]{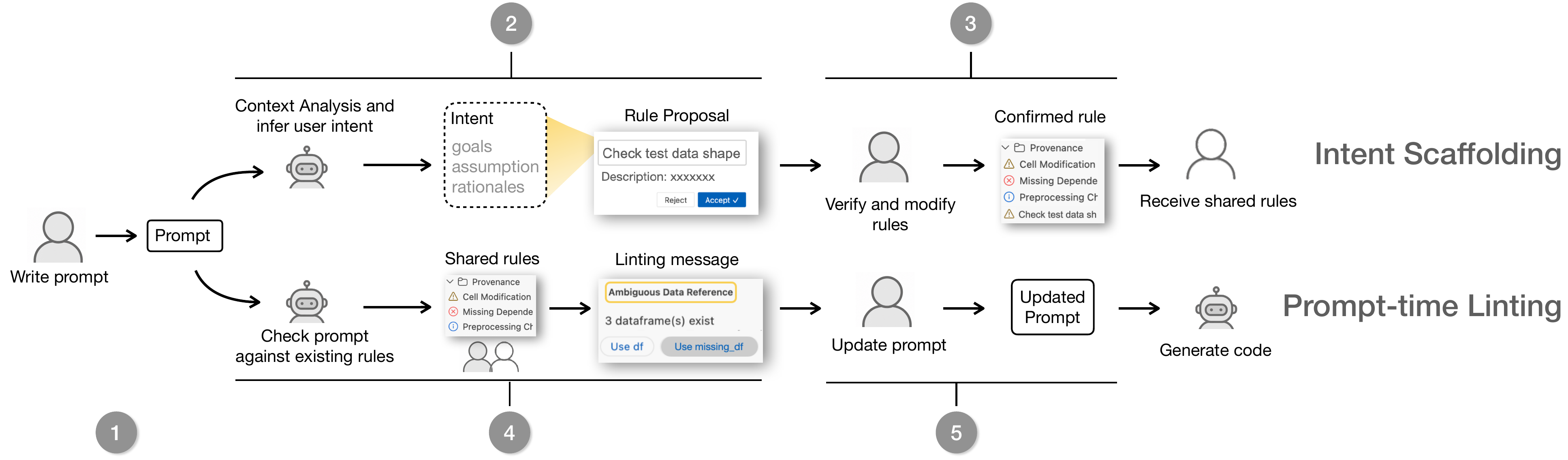}
    \caption{
    Overview of how our mechanisms work. When Devin submits a prompt (1), the rule-proposal agent analyzes the shared context and prompt, infers user intent, and proposes a rule for Devin to review, which is a coordination norm capturing Devin's intent (2). Once Devin verifies the rule proposal, it becomes a confirmed rule and is shared with other teammates, such as Maya (3). In parallel with (2), the rule-checking agent checks the current prompt against all existing rules. If the rule-checking agent detects potential conflicts with collaborators’ work, it will call the linting agent to generate a linting message that highlights prior analytic intent, relevant context, and suggested prompt revisions (4). Guided by this linting feedback, Devin updates the prompt for final code generation (5).
    }
    \Description{}
  \label{fig:teaser}
\end{teaserfigure}


\maketitle

\input{_body.tex}

\begin{acks}
This work was made possible by the Natural Sciences and Engineering Research Council of Canada (NSERC) Discovery Grant \#RGPIN-2025-01552, NSERC Discovery Grant \#RGPIN-2020-03966, NSERC Research, Technology, and Instruments Grant \#RTI-2026-00134, the Ontario Early Researcher Award (ERA) \#ER24-18-222, and the Canada Foundation for Innovation (CFI) John R. Evans Leaders Fund (JELF) \#42371.
\end{acks}

\bibliographystyle{ACM-Reference-Format}
\bibliography{_references.bib}


\appendix
\input{sections/appendix}


\end{document}

%% file: _setup.tex
\usepackage{exii-macros}

\usepackage{booktabs}
\usepackage{array}
\usepackage{subcaption}
\usepackage{fontawesome}
\usepackage{multirow}
\usepackage{graphicx}
\usepackage{lipsum}
\usepackage{svg}
\usepackage{makecell}
\usepackage{float}
\usepackage{cleveref}
\usepackage{xspace}
\usepackage{stfloats}
\usepackage{enumitem}

\definecolor{warm}{RGB}{235, 246, 249}

\usepackage{graphicx} 
\usepackage{array} 

\newcommand{\maxwidthbox}[1]{%
  \resizebox{\ifdim\width>\textwidth\textwidth\else\width\fi}{!}{#1}%
}

\definecolor{codecolor}{RGB}{239,239,237}
\definecolor{codetextcolor}{RGB}{165,0,0}

\definecolor{myblue}{RGB}{211, 231, 255}
\definecolor{mypurple}{RGB}{220, 198, 250}
\definecolor{myred}{RGB}{224, 182, 176}
\definecolor{termblue}{RGB}{0, 102, 204}

\definecolor{themered_E}{RGB}{236, 91, 75}
\definecolor{themeyellow_CS}{RGB}{243, 173, 60}
\definecolor{themegreen_VP}{RGB}{129, 165, 80}
\definecolor{themeblue_AP}{RGB}{131, 171, 249}
\definecolor{themepurple_CC}{RGB}{115, 73, 119}
\definecolor{themebrown_TA}{RGB}{159, 75, 27}

\newcommand{\sys}[0]{\textsc{IntentLint}\xspace}

\showrevisions{REVISIONGREEN}

\makeatletter
\g@addto@macro\normalsize{%
  \setlength\abovedisplayshortskip{-9pt}
  \setlength\belowdisplayshortskip{3pt}
}
\makeatother

\def\markup{0}

\if\markup1
\newcommand{\rv}[1]{{\leavevmode\color{blue}#1}}
\else
\newcommand{\rv}[1]{#1}
\renewcommand{\st}[1]{}
\newcommand{\sout}[1]{}
\fi

\usepackage{listings}

\lstdefinestyle{mystyle}{
    basicstyle=\ttfamily\small,
    columns=fullflexible,
    showstringspaces=false,
    frame=tb,
    numbers=left,
    numbersep=10pt,
    numberstyle=\tiny\color{black},
    stringstyle=\color{blue},
    commentstyle=\color{green!50!black},
    morestring=[b]",              
    morecomment=[l]{//}           
}

%% file: _body.tex

\input{sections/introduction-V1}
\input{sections/related-work}
\input{sections/formative-v2}
\input{sections/system-v2}
\input{sections/study}
\input{sections/discussion}
\input{sections/conclusion}

%% file: sections/introduction-V1.tex
\section{Introduction}
Data analysis is fundamentally collaborative: analysts must coordinate and maintain a shared understanding of what has been done, why it was done, and what should happen next~\cite{zhang2020data,dourish1992awareness,robinson2008collaborative}. In practice, this shared understanding is often fragile because the artifacts meant to capture it, such as computational notebooks and external documents, often become incomplete, outdated, or difficult to interpret as analyses evolve~\cite{wang2022documentation, Rogers:AutoMLTrace:2023, koesten2019collaborative}. Even when they are of high quality, collaborators must still expend significant effort to interpret them and infer collaborators' intent behind prior decisions~\cite{wang2020callisto, wiseman2017challenges}. As large language models (LLMs) are increasingly used to support data analysis, they have shown promise in assisting with code generation, explanation, and conversational interaction with data~\cite{tian2024chartgpt, maddigan2023chat2vis}. These systems suggest that LLMs could help reduce the burden of interpreting collaborators' work by synthesizing documentation, summarizing analyses, and generating explanations within ongoing workflows.


However, these advances do not fully address the demands of coordination and maintaining shared understanding in human-AI collaborative analysis. As LLMs become active participants in analytic workflows, coordination is required not only among human collaborators, but also between people and AI agents, and across multiple AI agents~\cite{chopra2023conversational}. This shift introduces new challenges, including supporting shared mental models, enabling inspection of agent reasoning, and maintaining coherence across agents~\cite{Zamfiresco:BadPrompt:2023}. Yet existing approaches primarily support artifact production and retrospective documentation~\cite{zhou2025xavier, wang2022documentation, tian2024chartgpt}, offering limited support for these coordination challenges in human-AI collaborative analysis.

To better understand these challenges, we conducted a formative study with five data analysts performing data analysis tasks. We found that analysts struggled to externalize analytic intent in ways legible to both collaborators and AI agents; undocumented assumptions propagated silently into AI-generated code; and conflicts arising from misaligned decisions were rarely surfaced before causing downstream inconsistencies. These breakdowns suggest a need for mechanisms that (1) make intent explicit and shareable, and (2) proactively detect misalignment during interaction.

\textbf{We introduce a rule-based coordination layer for AI-assisted collaborative notebooks with two interaction mechanisms, intent scaffolding and prompt-time linting, to encode intent as persistent shared rules and enable early issue detection.}
Together, these mechanisms form a feedback loop between intent articulation and action: making reasoning explicit and reintroducing it at critical moments to guide ongoing analysis.
We demonstrate them through a proof-of-concept system, \sys{}. \sys{} analyzes computational notebooks to extract context about code dependencies, dataflow, and prior analytic results. From this context and user prompts, it infers likely goals, assumptions, and decision rationales, presenting them as structured, editable rules for analysts' review. When a user issues a new prompt, \sys{} checks it against active rules and flags potential conflicts before code generation, helping steer agent behavior toward outputs consistent with the team’s analytic goals.

We evaluated \sys{} with 16 participants with data analysis backgrounds. Results show that \sys{} helped participants better understand collaborators’ activities and intent, reflect deeper on their own analytical choices, and identify misalignments earlier. Analytic intent scaffolding encouraged participants to make their strategies explicit and revisit them over time, while prompt-time linting surfaced conflicts between current prompts and previously established goals or constraints. Together, these results suggest that intent scaffolding and prompt-time linting can strengthen shared understanding in human-AI collaborative data analysis.

Our research provides the following contributions:
\begin{itemize}
    \item \textbf{A formative study} that identifies breakdowns in analytic intent and knowledge sharing across multi-human, multi-agent data science workflows, revealing design opportunities for interaction techniques that support collaboration and alignment.
    \item  \textbf{A proof-of-concept system} that operationalizes analytic intent scaffolding and prompt-time linting, enabling analysts and AI agents to align on goals, assumptions, and decisions in data analysis.
    \item \textbf{A user study} that demonstrates how real-time scaffolding and linting improve usability, collaboration, and understanding of shared analytic intent, and provides actionable design implications for future human–AI collaborative systems.
\end{itemize}

Despite advances in AI-assisted analysis, our research shows that collaborative data analysis workflows often break down when analytic intent is implicit or when coordination is unclear. We demonstrate how structuring collaboration through intent scaffolding and prompt-time linting enables analysts and AI agents to build on prior analyses and coordinate more effectively.

%% file: sections/related-work.tex
\section{Related Work}
We review prior work on collaborative data analysis and AI-assisted workflows, focusing on shared understanding and intent alignment.

\subsection{Collaborative Data Analysis}
Maintaining common ground is a fundamental challenge in collaborative data analysis, as intermediate artifacts are often exploratory and difficult to interpret~\cite{heer2007design, Beth2017Exploring}. 
Documentation plays a key role in sustaining shared understanding~\cite{vance2022creating}, and prior systems have explored varied designs to support this. 
Systems like Callisto, Verdant, and KTGraph support sensemaking by linking discussions to artifacts, visualizing analysis histories, and externalizing reasoning through annotations and provenance tracking~\cite{wang2020callisto,kery2019towards,zhao2017supporting}.
Other work focuses on cross-role knowledge transfer, such as ZIva, which facilitates communication of domain expertise to data scientists~\cite{park2021facilitating}.
However, these systems primarily focus on human–human collaboration. The introduction of LLMs fundamentally complicates knowledge sharing, as systems must also account for how human knowledge is communicated to AI agents.

Recent work explores knowledge sharing in LLM-mediated environments. Tools like CoAIcoder, SemanticCommit, and CoPrompt leverage context, detect semantic conflicts, and support reuse of prompting strategies, while studies note that prompts are often not preserved, limiting reproducibility and collaboration \cite{gao2023coaicoder,vaithilingam2025semantic,feng2024coprompt,haase2025towards}.
Other systems, such as Respark and PolicyPad, focus on transferring human knowledge to guide LLM behavior \cite{tian2025respark,feng2025policypad}.
However, these systems either overlook the collaborative dynamics of multi-human settings or treat knowledge mainly as constraints for AI agents rather than shared artifacts for all participants. 
Building on this gap, our work explores knowledge sharing across multi-human and agent workflows, where analytical knowledge is jointly constructed, negotiated, and reused.

\subsection{Intent alignment in AI-assisted data analysis}
LLM agents with different roles have been applied across the data science workflow~\cite{rahman2025llm}.
from translating natural-language queries into visualizations (e.g., Chat2VIS, ChartGPT) to supporting higher-level, concept-driven analysis authoring (e.g., Data Formulator) and decomposing complex analytical intents into structured subtasks (e.g., DocWrangler, LightVA)~\cite{rahman2025llm, maddigan2023chat2vis, tian2024chartgpt, wang2023data, zhao2024lightva}. Beyond individual interactions, recent systems investigate LLM-driven analysis workflows. InsightPilot autonomously generates sequences of analysis steps and insights \cite{ma2023insightpilot}, while DataNarrative employs multi-agent architectures to generate and verify analytical narratives \cite{islam2024datanarrative}. Data Director and similar systems extend this paradigm to multimodal outputs, such as automated data videos \cite{shen2024data}. However, these approaches often limit user control and provide little support for expressing user intent throughout the workflow.

\rv{Other work focuses on improving user awareness and interaction with AI-generated outputs. Prior systems support intent alignment by enabling users to steer and verify AI analyses through editable plans and intermediate representations \cite{kazemitabaar2024improving, xie2024waitgpt}, improving alignment between user prompts and generated scripts \cite{zhu2025visegpt}, and interactively resolving ambiguous user intent \cite{gao2015datatone, mu2024clarifygpt}. }
Related systems also improve verification by exposing richer data context, such as Xavier, which integrates dataset semantics into code suggestions \cite{zhou2025xavier}, and Amplio, which helps analysts explore underrepresented regions of datasets \cite{yeh2025exploring}. At a broader level, studies of agent configuration practices show that developers tend to encode functional instructions but rarely specify higher-level constraints such as intent, preferences, or quality criteria \cite{chatlatanagulchai2025agent, jiang2025empirical}. Benchmarks such as BLADE show that LLM agents struggle with open-ended analytical tasks, producing shallow and low-diversity analyses \cite{gu2024blade}. 
\rv{While these approaches establish effective mechanisms for intent alignment in single-user, single-session analysis, our work addresses the additional coordination challenges of maintaining shared, evolving intent across multiple humans and AI agents.}

Emerging research highlights the importance of making AI reasoning and coordination more transparent. Systems such as Cocoa demonstrate that shared, inspectable plans improve human–AI collaboration \cite{feng2025cocoa}, while collaborative environments integrating visible agent roles and actions reshape coordination practices \cite{lehmann2025collaborative}. Similarly, Narrative Scaffolding emphasizes externalizing reasoning processes to support more robust and defensible analysis \cite{huang2025narrative}. However, intent is still largely treated as a static input or plan rather than a shared, evolving construct. Our work builds on this gap by supporting intent alignment as an ongoing, collaborative process across both human and AI participants.

%% file: sections/formative-v2.tex
\section{Formative Study}\label{sec:formative}
\input{tables/challenge}

We conducted a formative study to better understand the challenges in collaborative data analysis involving multiple people and an AI coding assistant. Grounded in issues identified in prior literature (\Cref{tab:challenges}), we employ a Jupyter notebook as a technology probe to situate, re-examine, and extend these challenges in the context of contemporary AI code-assistant tools. Our study focused on asynchronous workflows, where analysts must interpret changes after they occur, whether made by themselves, other people, or AI assistants~\cite{Tory:DataVoice:2023}. This setting presents realistic and unique challenges for coordination and maintaining shared knowledge.

\vspace{-1mm}
\subsection{Technology Probe Development}\label{sec:formative-tech-probe}
The study of collaborative data analysis, both with and without automated support (e.g., AutoML~\cite{Wang:AutoML-needs:2019,Rogers:AutoMLTrace:2023} or AI assistants~\cite{ma2023insightpilot,maddigan2023chat2vis}), is well established. Rather than re-eliciting known challenges through self-report, which risks abstract and incomplete accounts, we build on prior work by employing a computational notebook as a technology probe that instantiates these challenges as concrete scenarios, enabling participants to engage with them in realistic workflows.

\vspace{-1mm}
\subsubsection{Defining Collaborative Data Analysis Challenges}\label{sec:formative-challenges}
As part of our formative study, we constructed a set of 15 challenges in collaborative data analysis from 25 state-of-the-art papers (\Cref{tab:challenges}). We focused on issues related to knowledge and context sharing among people. Papers were selected from HCI, CSCW, and software engineering based on their influence and relevance to computational notebooks - the workhorse of data analysis.  Using a light-weight open coding, we identified recurring problem descriptions and design observations, and iteratively consolidated them into 15 challenges organized into four categories. As we analyzed additional papers, few new themes emerged, suggesting sufficient coverage for our purposes. This synthesis is not intended to be exhaustive, but rather to provide a representative baseline that grounds our formative study; participants were able to critique, extend, and introduce additional challenges during the study.

\vspace{-1mm}
\subsubsection{Construction a Computational Notebook Technology Probe}\label{sec:formative-probe}
Each scenario was implemented as a realistic but synthetic code segment within a single Jupyter notebook, grounded in the Ames Housing dataset~\cite{kaggleAmesHousing} to maintain a consistent analytical context.  The notebook was pre-executed to ensure uniform presentation, with each scenario depicting a plausible failure state, such as a silently overwritten variable or an undocumented preprocessing step. On the study, Participants could freely edit and re-execute cells. The full notebook is included in the supplementary materials.

\vspace{-2mm}
\subsection{Procedure, Participants, \& Data Collection }\label{sec:formative-procedure}
Participants provided consent at the beginning of each session. Each participant received a unique instance of the computational notebook probe. After taking a few minutes to familiarize themselves with the notebook content, they were then guided through 15 brief scenarios illustrating the challenges summarized in \Cref{tab:challenges}. In each scenario, participants were shown the actions of two other human users (User A and User B) and asked to reflect on their understanding of these actions from the perspective of each user, gaps and challenges in understanding the user's actions, and how they could be supported by AI assistants or system affordances. We encouraged participants to think-aloud~\cite{lewis1982using} to capture their thought processes. Participants were also asked to reflect on their own experiences encountering these challenges and shared expectations for how LLMs might assist in addressing them. At the end of the session, participants could further elaborate on the challenges presented or propose additional ones. Sessions lasted between 60 and 80 minutes.
The study was approved by our institution's Ethics Board. Participants were remunerated \$30/hr via a digital gift card, with rates adjusted according to the time spent.  

\vspace{-1mm}
\subsubsection{Participant Recruitment \& Demographics}\label{sec:formative-participants}
We recruited five participants (2 males, 3 females; ages 23 - 26) through purposive sampling~\cite{etikan2016comparison}, targeting data professionals who routinely collaborate with others in analysis workflows and use AI coding assistants.  Participants were screened on the basis of years of professional data analysis experience, frequency of conducting data analyses, and self-reported familiarity with AI coding assistants (interpreted broadly as using formal coding agents to chatbot interactions asking for coding advice, etc.). All recruited participants had more than three years of data analysis experience, and regularly used LLM-driven tools (See appendix for full demographic details). 

\vspace{-1mm}
\subsubsection{Data Collection and Analysis}\label{sec:formative-data-analysis}
We collected modifications that participants made to each notebook instance, recordings of their sessions, and any notes made by the study administrator. Audio was transcribed and analyzed using an inductive thematic analysis, seeking to surface insights that informed design goals, rather than strict inter-coder reliability~\cite{terry2017thematic,Braun01012006}.

\vspace{-1mm}
\subsection{Findings}\label{sec:formative-findings}

Data analysts collaborating on multi-author notebooks face coordination, context, and reasoning challenges that are amplified when AI assistants are involved.  Although some challenges resemble those arising from human collaborators, the absence of social dynamics in AI changes how participants experience and respond to these interactions. Participants reflected on the persistence of these challenges over their careers and their expectations toward how AI agents should respond to support ongoing, asynchronous collaboration between people.

\subsubsection{Persistent challenges with collaborative data analysis}\label{sec:formative-finding-challenges}
Across participants, we observed persistent challenges in collaborative data analysis that span the categories identified in prior work (C1–C15). 

\vspace{-1.5mm}
\paragraph{Code management \& technical debt (C1-C3)}
Asynchronous updates by people or coding agents contributed to challenges in managing changes in the notebook, particularly when there were execution errors (C1). Participants reported that this problem was exacerbated by coding agents because, unlike humans, it was often difficult to seek clarification on their intentions behind changes and resulting errors. Of particular concern was when \rv{coding} agents acted on their own accord, by overwriting user code to introduce unstable dependencies (C2), modifying functions and variables (C3), and otherwise failing to consider human objectives in the existing code block. While both human and AI-driven modifications can introduce technical debt, participants reflect that AI technical debt, particularly verifying its outputs, was more disruptive and required active mitigation.

\vspace{-1mm}
\paragraph{Analytical rigor \& methodology (C4–C6)}
Code-management and issues of technical debt ultimately contributed to questions about the overall rigor of the data analysis. Separating between human-generated and AI-generated code was especially concerning, with participants raising the propensity of coding agents to hallucinate, or confidently provide inappropriate suggestions (e.g., algorithmic choice). As P1 mentioned, \pquote{AI generations [sic] cannot be perfect, so this uncertainty needs to be portrayed to the user.} This often made it difficult for participants to understand the outputs of analysis algorithms (C4), the logic of the overall analysis (C6), and the rationale for the choices that have been made (C5). Participants reflected that when working with people, discussion played a key role in resolving these ambiguities. However, the same nuanced discussion was more limited with coding agents. Additionally, P2 expressed a preference for AI to proactively ask clarification questions rather than producing inferences based on incomplete information. 

\vspace{-1mm}
\paragraph{Data governance \& version control (C7-C10)}
As datasets grow in size and complexity, participants expressed doubt that coding agents could (1) adequately understand these data (C7), particularly in pre-processing steps, and (2) reliably record changes to the data state (C8) along with the methodological evolution of the analysis workflow.
They were especially concerned with AI's ability to reliably detect issues in the workflows (C9, C10), noting that \pquote{human collaborators can usually infer this is probably a mistake, but AI may assume humans always make the right decisions.}{P1}. Participants also expressed differing preferences regarding how AI should intervene when issues are detected. Some preferred that AI provide suggestions along with rationales for code fixes, while others favored automated fixes for straightforward issues.

\vspace{-1mm}
\subsubsection{Documentation \& reproducibility (C11-C15)}
Finally, participants highlighted that LLMs are limited in supporting documentation and reproducibility. Data analysis is iterative updating to new findings and observations~\cite{kery2018interactions}. This makes documenting assumptions particularly important (C11-C13). These add to technical debt by adding to the burden of \pquote{reading tons of AI generation}{P5}. However, these challenges went both ways. P1 noted that when collaborators do not explicitly document their reasoning and decisions, AI suggestions about which directions to follow can create additional risks. Moreover, participants expressed skepticism that coding agents, unlike people, could reliably interpret intermediate results (C14) or evolving analysis logic (C15). At the same time, externalizing higher-order analytical thinking and reflection, whether by humans or AI, introduces significant cognitive overhead.

\vspace{-1mm}
\subsubsection{New challenges in HAI collaboration data analysis}\label{sec:formative-finding-challenges}
Participants also expressed new challenges for how they wished to work collaboratively with fellow analysts and AI-agents. These reflections are rooted in those defined in \Cref{tab:challenges}, and represent how collaborative human-AI data analysis should evolve.

Participants repeatedly highlighted the importance of shared contextual information in collaborative notebooks, including: factual notebook details, stated and inferred analytic intent of users and AI assistants, and suggested next steps for analysis. 
\rv{While prior research highlights over-reliance on confident AI outputs, our study contextualizes this concern within shared notebooks where participants emphasized that facts should be presented before inferred content and clearly distinguished (P2).} Preferences varied in the amount of context shown: some participants wanted a detailed view, while others favored a lighter presentation focused on facts and relevant background. Most participants also wanted options suggesting solutions or prompt improvements, both to reduce effort and to verify whether the model understood user intent correctly (P5). To avoid disruptions, participants voiced that timing was important, particularly for AI agents: real-time suggestions were helpful for immediate tasks, but non-urgent guidance should be surfaced at natural milestones, such as when committing changes (P3), to avoid disrupting exploratory work. Subtle visual cues that integrate with existing workflows were also critical for effective presentation.

\subsection{Design Guidelines}
The challenges of our formative study motivate the following design guidelines for aligning shared human and AI knowledge and context in collaborative data analysis. 

\begin{itemize}
    \item \textbf{DG1: Externalize analytic intent into a shared human–AI representation.} Many challenges arise when analytic intent remains implicit (C4-7,10-15). Coding conventions address some issues, but most are context-dependent, requiring semantic understanding of workflows and collaborators’ intentions. Systems should externalize analytic intent into a shared representation that both humans and LLMs can operate on, supporting constraints from static rules to context-aware checks. 
    \item \textbf{DG2: Reduce cognitive burden through context-aware guidance.} 
    Coordination in data analysis requires analysts to track the notebook state and prior decisions as they progress (C1-3,7-9,14). Systems should reduce this burden by providing context-aware guidance, surfacing relevant context at the point of action so users do not have to retrieve prior context themselves.
    \item \textbf{DG3: Support iterative refinement of collaboration norms.} 
    As analysis unfolds, collaboration norms, such as how data should be handled and which assumptions should guide the work, often need to change as well (C2,5-9,12,13,15). Systems should support iterative refinement of these norms so shared expectations can evolve with the analysis.
\end{itemize}

%% file: tables/challenge.tex
\begin{table*}[t]
\footnotesize
\centering
\maxwidthbox{

\begin{tabular}[]{p{3.5cm}p{8cm}p{8cm}p{1.5cm}}
\toprule
\textbf{Category} & \textbf{Challenges in human collaborative data analysis} & \textbf{Challenges in human-AI team collaboration} & \textbf{Reference} \\
\midrule
\multirow[t]{3}{*}{Code management \& technical debt} 
    & C1: Non-executed code cells with only code comments result in confusion & Confuse agents about the cell’s effect and lead to inappropriate code generation &\cite{pimentel2021understanding} \\
    & C2:  Modification of code cells used by others (unstable cell dependencies)  & Agent modifies code without explicit notification to humans &\cite{li2020resolving, pimentel2021understanding} \\
    & C3: Unintentional function and variable redefinitions &  Agent redefine variables without explicit notification to humans &\cite{wang2024don, pimentel2021understanding}\\
\midrule
\multirow[t]{3}{*}{Analytical rigor \& methodology} 
    & C4: Difficulty interpreting algorithmic outputs (e.g., correlation coefficients) & Distinguish human work and AI generation &\cite{passi2018trust, ramasamy2023visualising} \\
    & C5: Subjective pattern identification without documented criteria & Agent infer pattern definition without surfacing uncertainty &\cite{alspaugh2018futzing, battle2019characterizing, kandel2012profiler, koonchanok2021data} \\
    & C6: Unclear methodology for pattern validation &  Agent infer methodology without surfacing uncertainty &\cite{zgraggen2018investigating, zhao2017controlling, kale2023evm, koonchanok2023visual}\\
\midrule
\multirow[t]{4}{*}{Data governance \& version control} 
    & C7: Assuming data structure without verification or documentation  & Reliability in understanding large datasets without explicit guidance &\cite{kandel2012enterprise, chattopadhyay2020s, kandel2012profiler} \\
    & C8: In-place data mutations without version tracking  & \rv{Confuse agents about dataset state, causing inconsistent analyses and conflicts} &\cite{Head2019Managing}\\
    & C9: Neglected data cleaning steps leading to unreproducible or misleading results & Struggle to interpret if issues are from data or preprocessing &\cite{chicco2022eleven}\\
    & C10: Ambiguous data source  & Agent infer data source without surfacing uncertainty &\cite{kandel2012enterprise, Wang2019How}\\
\midrule
\multirow[t]{5}{*}{Documentation \& reproducibility} 
    & C11: Undocumented failed analysis attempts or exploratory paths & Cause agents to repeat failed analyses or miss context & \cite{kery2018interactions, kery2018story, zhang2020data}\\
    & C12: Lack of documentation for discretionary pre-processing decisions  & More effort spent reading generated explanations that offer little value & \cite{passi2018trust, muller2019data}\\
    & C13: Lack of documentation for data transformations  & Agent infer data source without surfacing uncertainty & \cite{kandel2012enterprise, rule2018exploration}\\
    & C14: Difficulty understanding how results (charts, tables) were produced  & Agent provides potentially incorrect inferences & \cite{Head2019Managing, kery2018story} \\
    & C15: Lack of documentation for statistical assumptions and methodology & Agents apply inconsistent methods and generate conflicting conclusions &\cite{rule2018exploration, kandel2012enterprise} \\
\bottomrule
\end{tabular}
}
\caption{Challenges in collaborative data analysis identified from 25 prior studies\vspace{-3mm}}
\label{tab:challenges}
\end{table*}

%% file: sections/system-v2.tex
\section{\sys{}} \label{sec:design}
\begin{figure}
    \centering
    \includegraphics[width=1\linewidth]{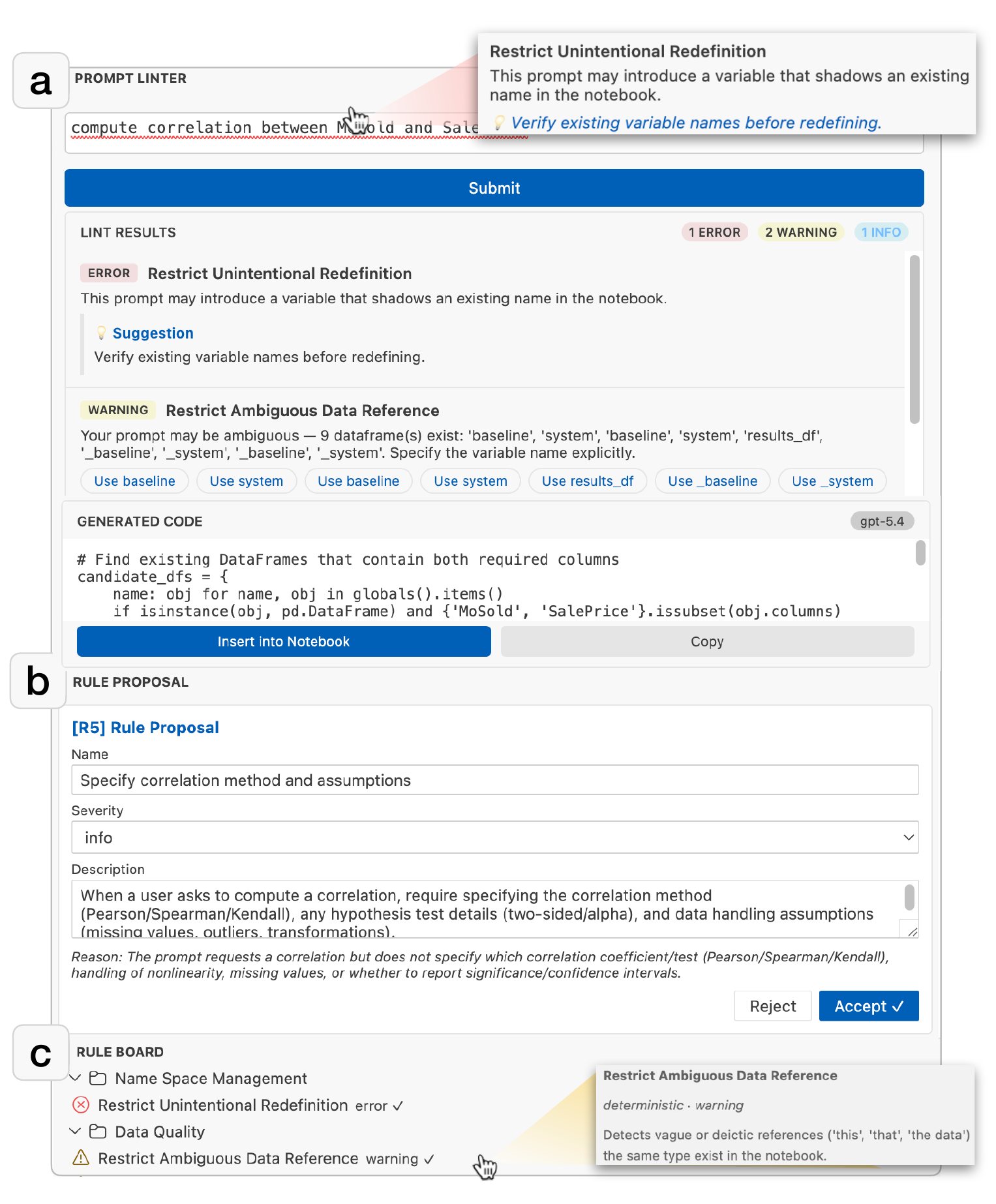}
    \caption{The \sys{} interface features three panels: Prompt Linter(a), Rule Proposals(b), and Rule Board(c). The prompt linter displays linting results for the current user prompt, along with explanations and suggested improvements. The Rule Proposals panel allows users to review and edit proposed rules before adding them to the system. The Rule Board displays all active and inactive rules, which users can view and manage.}
    \label{fig:UI2}
\end{figure}

We developed \sys{}, a VSCode extension that helps data analysts maintain shared knowledge among humans and AI \rv{coding} agents in collaborative data analysis. Guided by findings from our formative study, we propose and operationalize \rv{ a rule-based coordination layer for AI-assisted collaborative notebooks, supported by}
two novel interaction mechanisms: \textbf{analytic intent scaffolding} and \textbf{prompt-time linting}. These mechanisms are enabled through shared rules, linked to collaborative challenges,  which facilitate coordination and alignment across human and AI collaborators \textbf{(DG3)}. The interface allows users to propose, review, and manage rules (Figure \ref{fig:UI2}) and connects with LLM-based tools such as GitHub Copilot and Claude Code.

\subsection{Core Interaction Mechanisms}\label{sec:design-core-interaction}
A key finding from our formative study was the burden of enforcing data, code, and analysis norms across people and AI \rv{coding} agents, leading to significant overhead and technical debt in managing evolving analyses. Participants valued documentation for orientation, but found it effortful to produce. To address these gaps, we proposed two interaction mechanisms:

\textbf{Analytic Intent Scaffolding} helps analysts externalize goals, assumptions, and decision rationales throughout data analysis \textbf{(DG1)}. Rather than relying on people or AI to infer intent implicitly, users iteratively clarify and refine their objectives as the analysis evolves. We argue that structured representations, rather than free-form comments or documentation, make analytic intent explicit, persistent, and verifiable. These representations can be generated by people or AI \rv{coding} agents, reducing cognitive burden for the former and making assumptions of the latter transparent. By surfacing goals, assumptions, and decisions clearly, they create a shared reference that can be revisited and built upon over time, supporting coordination, verification, and flexible engagement in collaborative analysis.

\textbf{Prompt-time linting} uses externalized analytic intent to provide timely, context-aware feedback during analysis \textbf{(DG2)}. Comparing actions against shared expectations helps ensure that human and AI collaborators remain aligned with goals and assumptions. This approach brings relevant context into the moment of action, reducing the need for analysts to manually track prior decisions. When ambiguities or potential conflicts arise, it should surface feedback that encourages reflection on how current actions relate to prior work. Considering alternative resolutions promotes accurate, aligned, and informed decision-making without requiring analysts to track prior decisions manually.

\textit{Together, these mechanisms form a feedback loop between intent articulation and action: Analytic Intent Scaffolding makes reasoning explicit and shareable, while Prompt-Time Linting reintroduces that intent at critical moments to guide ongoing analysis.}

\subsection{Rules: soft constraints in collaboration}\label{sec:design-rules}

To operationalize \textbf{Analytic Intent Scaffolding} and  \textbf{Prompt-time Linting} mechanisms, \sys{} implements shared rules that encode expectations about data, code, and analysis behavior. We chose rules as the core abstraction because they externalize tacit coordination norms into explicit, machine-readable forms without requiring formal specifications. Acting as boundary objects~\cite{star1989institutional}, they also provide concrete artifacts for negotiation: accepting, rejecting, or editing a rule becomes a coordination act that builds common ground \textbf{(DG3)}. By linking rules to human and AI actions, the system can surface timely, context-aware guidance \textbf{(DG2)}, detect potential conflicts, and maintain alignment across collaborators. 

\vspace{-1mm}
\subsubsection{Rule Types and Collaborative Analysis Challenges}\label{sec:design-rule-types-challenges}

\input{tables/rule-type}

Based on the prior literature and formative study findings, we derived a default set of rules that address the collaborative analysis challenges. These are shown in \Cref{tab:rule-type}, along with the challenges they aim to address. Users can modify, remove, or add to this core rule base (\S\ref{sec:design-rule-interactions}), and our user study directly assesses their utility as defaults. 

Each session is instantiated by the default set of rules. These rules can be modified, removed, or deactivated during a session, and new rules can also be proposed. Rules are contained as part of the notebook's rule board (\Cref{fig:UI2}), and persist after an analysis session to be used by other people or AI agents at a later time.

\vspace{-1mm}
\subsubsection{Rule Structure}\label{sec:design-rule-structure}
Rules follow a standardized template that is interpretable by both people and AI coding agents. This template provides a scaffold for expressing analytic intent, capturing assumptions, expected actions, and contextual dependencies in a structured form that can guide reflection and coordination across collaborators. An example template is below:

\begin{lstlisting}[style=mystyle]
{
  name: "Warn about Cell Modification Cascade"
  author: {User A}, // system or user ID
  level: 'info' | 'warn' | 'error',
  description: "Detects when a user's prompt is...",
  trigger condition: "The user's prompt...",
  context: "Commented cell xx...",
}
\end{lstlisting}

\noindent The components of the rule template are as follows:

\begin{itemize}
\item A \textit{\textbf{rule name}} describing the action that should be enforced.
\item The \textit{\textbf{author}} who constructed the rule, which can be a human or AI collaboration. This information is important for distinguishing between the actions of people and AI \rv{coding} agents \textbf{(DG1)}.
\item A \textit{\textbf{severity level}}, that encodes enforcement strength. Based on our formative study, we defined three levels: \textit{information}, \textit{warning}, and \textit{error}, which respectively surface relevant collaborator context, flag potentially important issues, and indicate likely violations that should be resolved before proceeding.  The level also determines how disruptive a triggered rule is, allowing intervention intensity to match the coordination need \textbf{(DG2)}.

\item A \textbf{\textit{Description}} makes the rule's reasoning transparent and verifiable. An LLM agent is designed to understand and trigger the rules based on the natural language descriptions \textbf{(DG1)}. 

\item \textbf{\textit{Trigger conditions}} that determine when a rule activates: rules fire only when relevant, not constantly. It distinguishes structurally detectable patterns (e.g., variable names, column references) from semantically rich conditions that require LLM interpretation, enabling the range of constraint types.

\item \textbf{\textit{Context}} that functions as a grounding field which records what notebook state (outputs, comments, prior code) caused the rule to be proposed. During early prototyping, we found that surfacing specific types of context (e.g., dataflow and code dependencies) improved the accuracy of LLM-based rule checking \textbf{(DG2)}. 
\end{itemize}

\subsection{Enforcing Rules via Prompt-Time Linting}\label{sec:design-rule-interactions}

\sys{} provides an interface for prompt-based interaction, similar to emerging AI-assisted coding tools, allowing users to modify computational notebook cells with AI support.  When users input a prompt, \sys{} automatically determines which rules should be applied, providing real-time linting information \textbf{(DG2)}. In \Cref{fig:UI2}, we show an example of this when a participant prompts an AI assistant to ``add correlation analysis of all the variables''. \sys{} warns that this prompt provides an ambiguous data reference (C10) and suggests variables from the analysis to revise the prompt to provide more information. In addition, \sys{} dynamically proposes a new rule, one specific to correlation analysis, that is informed by the triggering of the default and more generic R5 ('Require algorithmic/statistical test specification'; \Cref{tab:rule-type}); this new rule can be rejected or accepted by the user, growing the rule base without imposing additional overhead \textbf{(DG3)}.

\Cref{fig:teaser} illustrates this workflow. Bob enters a prompt, and \sys{} generates a structured rule that captures his analytic intent. Bob can review and accept this rule, allowing it to persist across analysis sessions as part of the shared context \textbf{(DG1)}. At the same time, \sys{} uses prompt-time linting to surface previously articulated intent from the whole team, including Bob, Alice, and other team members, warning Bob when his action may conflict with prior assumptions or decisions. Bob can choose to proceed, revise his prompt, or contribute new rules of his own, collectively shaping the shared analytical norms \textbf{(DG3)}.

\begin{figure}[h!]
    \centering
    \includegraphics[width=1\linewidth]{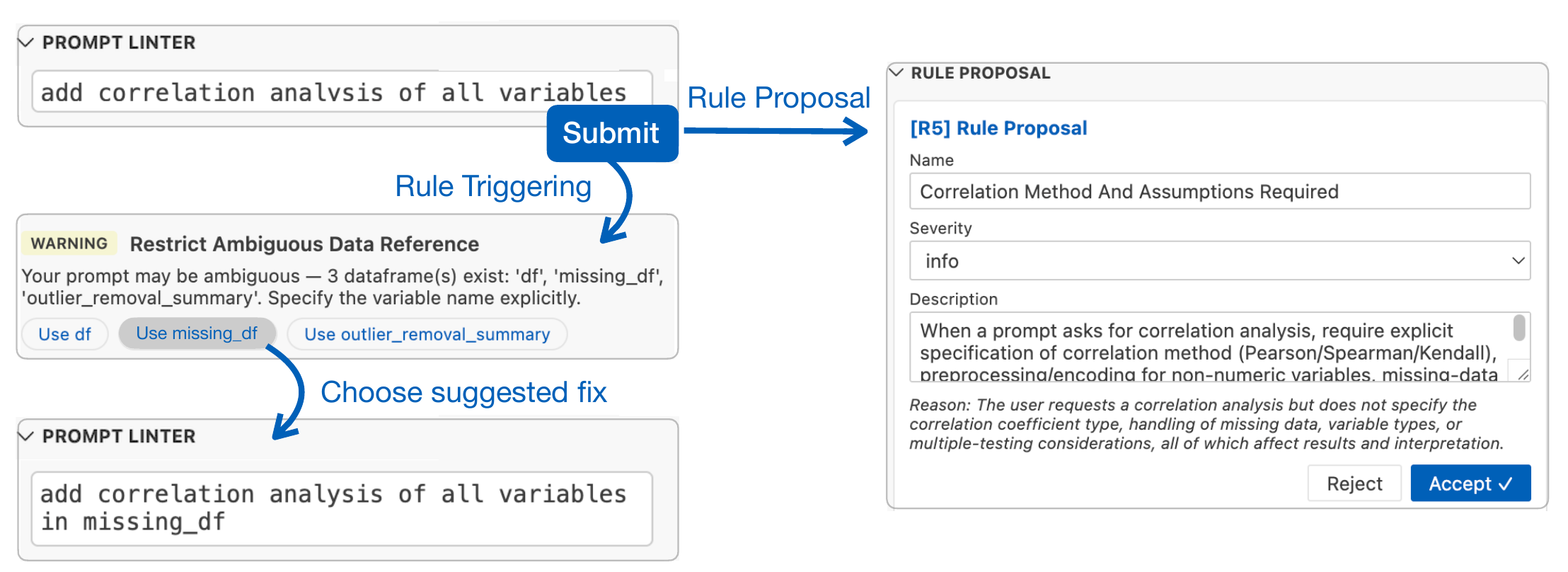}
    \caption{Rule-based Interactions: rule triggering and rule proposal. Upon prompt submission, \sys{} checks the prompt with shared rules and the current notebook context. A linting message with a suggested fix is displayed if a rule is triggered (left). A new rule is proposed based on the default rule and the current prompt (right).}
    \label{fig:interaction}
\end{figure}

\subsubsection{Prompt Time Linting Workflow.}\label{sec:design-workflow}
\sys{} uses a multi-agent back-end to dynamically apply, propose, trigger, and manage rules. \rv{Multiple} agents operate in parallel, but simultaneously return their results to the main \sys{} interface. In the appendix, we provide an architectural diagram as well as low-level technical details of this implementation. Here, we focus on conveying the high-level steps of this workflow. 

\vspace{-1mm}
\paragraph{Rule Triggering.} When a user submits a prompt, \sys{} evaluates it against the current notebook state and a set of existing rules \textbf{(DG2)}. A back-end `rule-checking' agent interprets each rule’s trigger condition to determine whether it applies to the prompt. Because these conditions are context-dependent, the agent dynamically assesses the alignment between the prompt and relevant rules. Multiple rules may be triggered and are aggregated in the \textit{Prompt Linter} panel, which displays the rule name, author, severity, and a contextual description of potential conflicts (\Cref{fig:interaction}).

\vspace{-1mm}
\paragraph{Suggested Prompt Updates.} When a rule is triggered, \sys{} uses its knowledge of the current notebook state, including code cell contents, cell indices, execution state, and execution order in addition to prompt and rules, to propose prompt revisions \textbf{(DG2)}. This is achieved through the `linting' agent that reconciles this information and proposes prompt modifications.  For example, in \Cref{fig:interaction}, \sys{} proposes specific variables from the notebook that could help reduce the ambiguity in the user's original prompt.  The suggested revision is surfaced automatically along with the linting message. Users can accept or decline these changes. 

\vspace{-1mm}
\paragraph{Propose new rules.} Finally, \sys{} can propose new rules based on those that are triggered, enabling more granular linting over time. A rule-proposal agent suggests candidate rules derived from existing ones, pre-populated with relevant context \textbf{(DG1)}. Users can review, refine, and add these rules to the rule board, allowing them to be applied in future interactions. This action is shown in the right-hand panel of \Cref{fig:interaction}.  The addition or modification of rules is propagated in real time to all collaborators, human and AI,  ensuring that the shared rule set remains consistent.

\vspace{-1mm}
\paragraph{Rule Management.} As rule sets expand, \sys{} can support personalized enforcement of different rules. Team members can directly modify each rule as the team’s analytical conventions evolve \textbf{(DG3)}. However, individual collaborators can also activate or deactivate any rule in their own workspace, allowing them to suppress rules that are irrelevant to their current analysis without affecting other collaborators. This distinction between team-level rule editing and individual-level activation supports shared conventions while preserving each collaborator’s autonomy.

\subsection{Implementation}\label{sec:design-imp}
The primary \sys{} client-side interface is implemented in Typescript. The backend is implemented using Python, using FastAPI to coordinate with the client-side UI,  and facilitates rule checking and prompt-time linting as well as manages the shared rule set. Additional details are in the appendix. 

%% file: tables/rule-type.tex
\begin{table*}
\footnotesize
\centering
\maxwidthbox{

\begin{tabular}[]{p{4cm}p{8cm}p{2cm}}
\toprule
\textbf{Category} & \textbf{Default Rule} &  {Resolved challenges} \\
\midrule
\multirow[t]{3}{*}{Namespace \& Definition Management} 
    & R1: Restrict unintentional variable and function redefinitions & C3 \\
    & R2: Require explicit data reference   & C7, C10 \\
\midrule
\multirow[t]{3}{*}{Dependency \& Side Effect Tracking} 
    & R3: Warn about cell modification cascade & C2 \\
    & R4: Require dependent cells to update  &C2\\
    & R6: Check in-place mutation versions  &C8\\
\midrule
\multirow[t]{4}{*}{Knowledge \& Context Preservation} 
    & R5: Require algorithmic/statistical test specification comprehension  & C4, C12, C14, C15\\
    & R7: Flag abandoned/alternative paths or operations  &C11\\
    & R8: Flag the impact of preprocessing on target data  & C9, C12, C13 \\
    & R9: Flag pattern/threshold definition for pattern comprehension and validation & C5, C6\\
\midrule
\multirow[t]{5}{*}{State \& Execution Consistency} 
    & R10: Restrict executing commented out cells &   C1\\
\bottomrule
\end{tabular}
}
\caption{Default rules instantiated in \sys{} and informed by the formative study challenges. Rules can be extended, modified, or removed by users, and new rules can be added or dynamically proposed by \sys{}. Rules persist in a notebook's workspace, allowing them to be asynchronously enforced over multiple analysis sessions and analysts. \vspace{-3mm} }
\label{tab:rule-type}
\end{table*}



%% file: sections/study.tex
\section{User Study}
We conducted a within-subjects study to evaluate the usability of \sys{}, by answering the following questions: (RQ1) How do participants perceive the effectiveness of rule proposing for scaffolding their intent expression? (RQ2) How do participants perceive prompt-time linting in supporting collaboration and exploration? (RQ3) Does the system \rv{with the rule-based coordination layer} change participants’ workflow? \rv{Our focus is not to evaluate if IntentLint produces better analytical outcomes, but whether it improves the collaborative analysis experience.}

The study was approved by a University Ethics Board. Sessions were an hour long. Participants were compensated at a rate of \$30/hr, with pay adjusted for sessions that exceeded this time. 



\subsection{Procedure, Participants, and Data Analysis}

\subsubsection{Study Conditions \& Tasks}
Our study included two conditions: a baseline and \sys{}. The baseline condition used GitHub Copilot \rv{(agent mode)}. Each condition was associated with a different task, instantiated through separate pre-populated computational notebooks designed to surface the collaborative challenges. These tasks and notebooks were co-designed with two experienced data analysts to simulate representative analysis workflows. 
\rv{The notebooks contain expert-authored inline comments and markdown cells, which were controlled across conditions.}
The order of conditions and tasks was counterbalanced in the study.

\vspace{0.5mm}
\noindent\textbf{Task 1:} Assess whether sale timing affects price by visualizing median sale price by month and year, and testing for differences across 2006–2010. Analyze the luxury segment by comparing Fireplaces and GarageArea, evaluating alternative thresholds (e.g., top 20\%), and updating preprocessing to remove near-zero variance features.

\vspace{0.5mm}
\noindent\textbf{Task 2:} Examine salary differences between attrition and non-attrition groups using visualization and statistical analysis. Complete preprocessing with one-hot encoding of key categorical variables, then analyze attrition patterns among high-performing employees across JobRole categories.

\subsubsection {Study Procedure}
Participants first provided informed consent, including permission for audio, video, and screen recording. They were then shown a demonstration of \sys{} and completed a set of practice questions to familiarize themselves with the interface. Participants were randomly assigned to begin with either the baseline or \sys{} condition. For each condition, they were given a pre-populated notebook corresponding to Task 1 or Task 2 and asked to build on existing work by completing additional analysis steps that extended prior results. Participants were encouraged to think aloud throughout the session. After 20 minutes, they switched to the other condition and task and repeated the process. At the end of the study, participants took part in a semi-structured interview, reflecting on how rule proposals and prompt-time linting influenced their approach to documenting analytic intent and building on collaborators’ work, as well as their expectations for integrating the system into their workflow and suggestions for improvement.

\subsubsection{Participant Recruitment and Demographics}
We recruited 16 participants ( 8 males, 8 females; ages 20 - 38) from a local university through mailing list, social media, and word-of-mouth. Potential participants were asked if they had prior experience with collaborative data analysis and AI tools such as GitHub Copilot and ChatGPT; they had to answer ``yes'' to both questions to participate. On a 5-point scale from 1-``a few times a year'' to 4-``daily'', they reported their frequency of use of AI tools for data analysis as $Mdn=3.5$ (See appendix for full demographic details). 

\subsubsection{Data Collection and Analysis}
\begin{figure*}[h]
    \centering
    \includegraphics[width=1\linewidth]{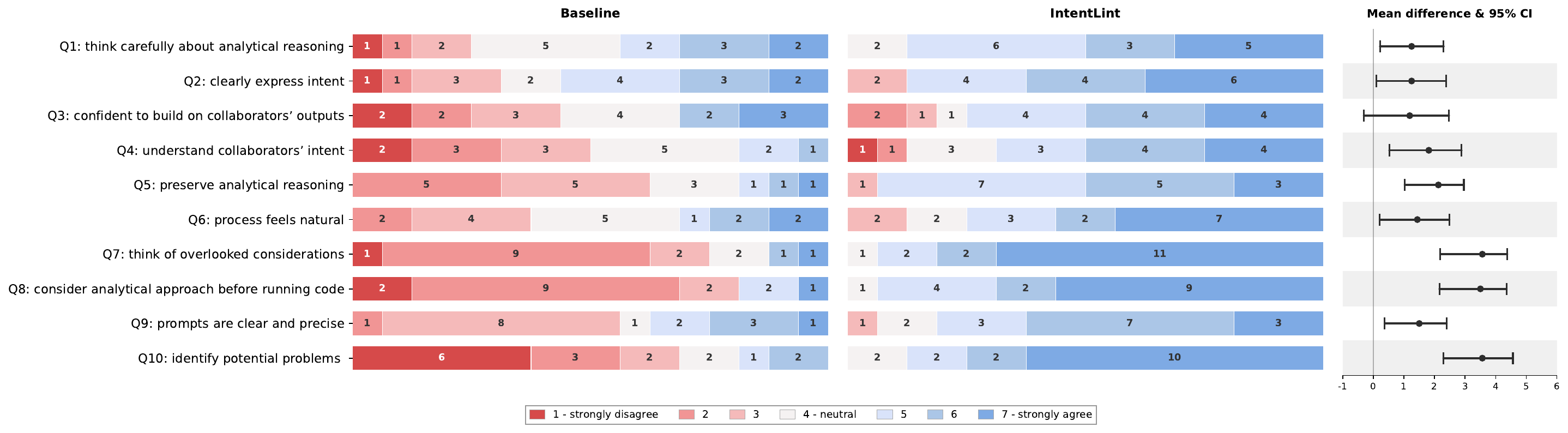}
    \caption{User perception of the utility of \textit{Baseline} and \sys{}, measured on self-defned 7-point Likert scales (Appendix \ref{appendix:self-likert}). Dots represent the mean differences of our system compared to the Baseline. Bars indicate the 95\% CI calculated using the studentized bootstrap method.}
    \label{fig:likert-chart}
\end{figure*}

Both the baseline and \sys{} automatically logged various types of events based on participants’ interactions during the study. At the conclusion of each condition, participants were asked to fill out a survey of ten questions (\Cref{fig:likert-chart}) on a seven-point Likert Scale (1= Strongly Disagree, 7=Strongly Agree)  about their experience, the standard system usability scale, and the NASA-TLX.  We summarize these results as descriptive statistics and compute statistically significant differences using a Wilcoxon signed-rank test.  Session audio was transcribed and analyzed using a reflexive thematic analysis~\cite{braun2019reflecting}. The goal of our analysis was to summarize overall participant impressions and how these support the quantitative findings.

\subsection{Findings}
\rv{\subsubsection{Rule-checking reliability} Because IntentLint's rule-checking interprets natural-language trigger conditions with LLM rather than applying deterministic pattern matching, we assessed how reliably it fires relative to human judgment. Following prior work that validates LLM-based classification in interactive systems by comparing system–human agreement against human–human agreement\cite{lam2025policy}, two human coders independently labeled 860 prompt-rule pairs from the user study log, judging for each pair whether the rule should fire given the prompt and notebook context. Inter-coder agreement was almost perfect (Cohen's $\kappa$ = 0.92). LLM–human agreement ranged 0.64–0.91 across rules (M = 0.86), indicating substantial agreement.}

\subsubsection{Overall rule usage and user behaviors}
\begin{figure}
    \centering
    \includegraphics[width=1\linewidth]{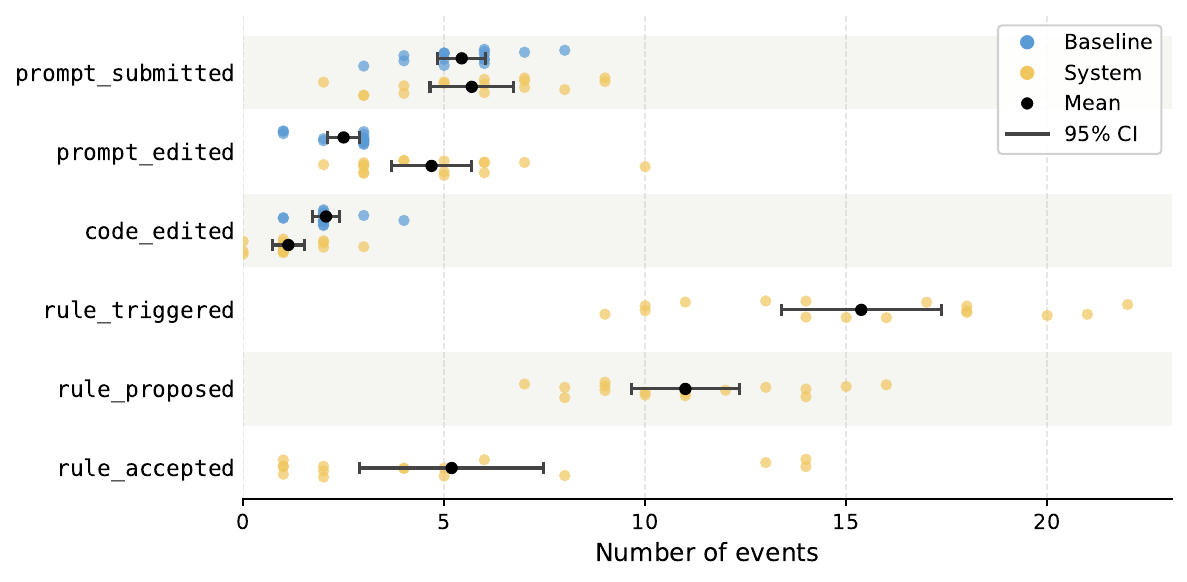}
    \caption{Distribution of system log events for 16 participants comparing the Baseline System and our system}
    \label{fig:event-counts}
\end{figure}

Participants actively engaged with all features provided in \sys{}. Across sessions, they interacted with linting messages, accepted auto-fix suggestions, edited prompts and generated code, reviewed and modified rule proposals, accepted or rejected proposed rules, and checked the rule board to monitor currently active rules (Figure \ref{fig:event-counts}). 

\vspace{-1mm}
\paragraph{Viewing linting suggestions and editing prompts}
Across all sessions, participants encountered an average of 22 linting messages per session. The linting messages covered all categories of rules and appeared in different levels of severity. The linting messages prompted participants to reconsider whether their prompts clearly expressed analytic intent and took collaborators' work into consideration. Consistent with this observation, we recorded substantially more prompt edits when participants used \sys{} ($N=75$) compared to the baseline condition ($N=40$). Participants also reported that their prompts were clearer and more precise with \sys{} ($Q9: Median\textsubscript{\sys{}}=6.0\ vs.\ Median\textsubscript{Baseline}=3.0, $ $p<.01$). They attributed this improvement to the actionable nature of the suggestions, which reduced the effort required to revise prompts: \pquote{The suggestions helped me to add specifications for the statistical test, which I am not familiar with.}{P15}. 


\vspace{-1mm}
\paragraph{Accepting proposed rules}
Overall, participants accepted a majority of the proposed rules (86 out of 157 distinct rules), suggesting that the system’s rule generation aligned well with collaboration concerns. However, we observed notable variation in acceptance behavior across individuals. 
Some participants adopted a permissive approach, accepting nearly all proposed rules, viewing them as ``safety nets'' that could catch potential issues without requiring immediate action. This indicates that our design of the rules and linting message is not intrusive.
Others were more selective, filtering out rules they considered unlikely to surface meaningful conflicts, as adding too many rules could overwhelm collaborators with constraints that distract from substantive analytic work.

\vspace{-1mm}
\paragraph{Editing rules}
Some participants modified the severity level of proposed rules, for example, changing them from ``info'' to ``warning'', to emphasize rules they considered more important. This adjustment reflected participants’ deliberation about which analytical practices should be more strongly enforced during collaboration. 
Participants also edited rule descriptions. By condensing the wording and removing potentially ambiguous or misleading phrases, they aimed to reduce the effort required for collaborators to understand the rules and to ensure the intended guidance was communicated clearly. Several participants also noted that clearer rule descriptions could help AI models better interpret the underlying analytical intent, reducing the risk of misalignment in subsequent prompts or generated analyses. These behaviors indicate that the structured rule representation prompted participants to actively reflect on how their analytic intent should be articulated and communicated to both collaborators and AI agents.

\vspace{-1mm}
\paragraph{System Usability \& Perceived Cognitive Load}
To measure the usability of \sys{}, we computed the SUS scores based on the UMUX-LITE~\cite{lewis2013umux}. Both \sys{} and Baseline have sufficiently reasonable usability scores, with an average of (M = 77.08) for \sys{} and (M = 68.75) for Baseline.
We also used NASA-TLX to measure participants’ perceptions of the cognitive workload of using the systems. The overall perceived workload, obtained by averaging all six raw NASA-TLX scores (with the ``Performance'' measure inverted), did not show a significant difference between \sys{} and Baseline ($\text{Mdn} = 2.67 $ vs $ 2.92, p=.28$).

\rv{\paragraph{Failure case analysis} We also identified several failure modes of rule triggering. One common issue arose from too strict variable-name constraints: variable names reused independently in two cells can trigger a cross-user conflict, even though the scopes never interacted (P5, P7). Another issue arose from coarse-grained dependency tracking: false dependency warnings were triggered by edits to unrelated lines, due to cell-level rather than line-level dependency tracking (P8, P10). These findings suggest that future conflict detection mechanisms should incorporate finer-grained code analysis and contextual awareness to reduce false positives.}

\subsubsection{Intent scaffolding encouraged deeper thinking}
The structured intent scaffolding prompted participants to reflect more carefully on their analytic strategies and assumptions. Participants reported thinking more carefully about analytical reasoning ($Q1: Mdn\textsubscript{I}=5.5\ vs.\ Mdn\textsubscript{B}=4.0, $ $p<.05$).
As shown in Figure \ref{fig:event-counts}, participants made more prompt edits to articulate intent using \sys{}. Several participants went beyond accepting the system's proposed rules and proactively composed rules themselves: \pquote{Some rules are not proposed by the system, but I want to define a rule when checking the code.}{P13} This behavior suggests that the scaffolding mechanism helped participants externalize their own analytic standards ($Q2: Mdn\textsubscript{I}=6.0\ vs.\ Mdn\textsubscript{B}=5.0, $ $p<.05$). 
Participants also reported becoming more deliberate in prompt formulation ($Q9: Mdn\textsubscript{I}=6.0\ vs.\ Mdn\textsubscript{B}=3.0, $ $p<.01$). For example, by including more detailed information about variables and statistical tests, reflecting a more explicit consideration of analysis requirements.

The linting messages also helped participants reason about the state of the collaborative analysis by surfacing issues and highlighting where attention was needed ($Q5: Mdn\textsubscript{I}=5.5\ vs.\ Mdn\textsubscript{B}=3.0, $ $p<.001$). As one participant noted, \pquote{It gave me the red flags in advance, so I know sections that we were supposed to alter}{P2}. In this way, \sys{} helped participants interpret prior work and decide where modifications were appropriate. Rule proposals further supported this reflection by surfacing analytic considerations participants might ignore. These interactions illustrate how structured rule feedback encouraged participants to reason about dependencies and methodological correctness in the analysis workflow.

Participants also compared the structured scaffolding with existing mechanisms for communicating intent, such as inline code comments ($Q4: Mdn\textsubscript{I}=5.5\ vs.\ Mdn\textsubscript{B}=3.5, $ $p<.005$). Several participants expressed skepticism toward free-text comments, particularly in collaborative environments where comments may be incomplete, outdated, or automatically generated. As one participant explained, \pquote{most comments are AI-generated anyway}{P3}. Instead, participants described the rule-based representations in \sys{} as a more dependable way to capture and communicate analytic expectations. These responses suggest that structured intent scaffolding can complement informal documentation by providing explicit, actionable representations of analytic intent that are more tightly integrated with the workflow.


\rv{Beyond supporting participants' interpretation of prior work, structured intent representations also influenced how they contributed to the shared workspace. Participants described adapting their own behavior with future collaborators in mind by making their analytic intent more explicit and curating what should become shared coordination knowledge. For example, P3 explained that they deliberately expressed their intent more clearly because \pquote{collaborators may not have time to thoroughly review the notebook}{P3}. Participants also became more selective about externalizing rules, rejecting later proposals that were already subsumed by broader existing rules to \pquote{avoid flooding the shared coordination space}{P11}. These observations suggest that intent scaffolding not only helps collaborators understand one another's work, but also encourages contributors to maintain concise and collaboration-oriented representations of analytic intent.}

\rv{We also asked participants to explain why they rejected proposed rules. Some participants explained that they tended to reject rules that represent one-time decisions rather than reusable coordination rules, or being subsumed by broader existing rules (P8, P13). Other participants found that some proposed rules are overly strict for the exploratory nature of EDA task they're working on, so they preferred to avoid unnecessarily constraining the team's workflow and to \pquote{avoid constraining too much for the whole team and annoying collaborators}{P7}. These responses suggest that effective rule recommendation should prioritize broadly applicable coordination practices while accounting for the flexibility and evolving nature of collaborative data analysis.}

\subsubsection{Prompt-time linting surfaces issues early}
All participants found prompt-time linting helpful for detecting potential errors and conflicts before adding code to the shared workspace ($Q10: Mdn\textsubscript{I}=7.0\ vs.\ Mdn\textsubscript{B}=2.0, $ $p<.001$). For instance, P15 highlighted the difference it made: \pquote{detecting the error was impossible before.}{P15}.
After resolving a variable conflict flagged by the extension, some participants reported increased trust in the generated code: \pquote{Now I have the confidence to say that the generated responses look good by just looking at these panels and warnings.}{P15}.
Beyond error detection, linting encouraged participants to review prompts and generated code, verify alignment with collaborators, and check data consistency with contextual explanations ($Q7: Mdn\textsubscript{I}=7.0\ vs.\ Mdn\textsubscript{B}=2.0, $ $p<.001$). As P4 noted: \pquote{Sometimes I just forget to check the data, and the agent gets back to me with context and explanations.}{P4}
It also reduces the effort required to resolve conflicts later in the workflow, as P6 mentioned: \pquote{It can reduce time needed for completing a pull-request.}{P6}
Although participants could deactivate or edit rules, only one chose to do so. Most perceived rules as non-intrusive and functioning as a safety net, reducing the need for deactivation. 


\subsubsection{Change of workflow: reflection on analytic intent and collaboration awareness}
Overall, participants demonstrated increased reflection on their analytic intent and whether their prompts clearly articulated that intent. 
This shift is consistent with the event logs (Figure \ref{fig:event-counts}), which show more prompt revisions when using \sys{} ($N = 75$ vs.\ $40$), and fewer direct code edits ($N = 18$ vs.\ $33$). 
These patterns indicate a shift in effort from code-focused activities toward the formulation and articulation of analytic intent.
Participants also dynamically adjusted rule severity over time, reflecting evolving confidence in their analyses and rule authoring. When initially unfamiliar with the dataset or collaborative context, participants tended to assign rules a lower (informational) severity. As their understanding deepened, they increasingly escalated rule severity to signal stronger expectations or constraints. For example, P14 explained their decision to upgrade a rule: \pquote{I think the choice of statistical tests is important, I want to warn my collaborators}{P14}. 

Participants also noted a shift in workload distribution. By offloading tasks such as data validation and conflict detection to \sys{}, they were able to focus more on iterating over analytic intent and decision-making. As P7 described, \pquote{The system can handle the potential conflicts for me}{P7}. This suggests that \sys{} supports a form of cognitive offloading that frees analysts to engage more deeply with higher-level reasoning.
In addition, participants identified specific moments in the workflow where the system was particularly valuable: during onboarding and when creating new data frames. These moments are characterized by uncertainty and a need for alignment, highlighting the role of \sys{} in establishing shared understanding early in analytic processes.
In addition, participants surfaced potential use cases and design considerations: Some described using the linter panel as a space for experimenting with and testing analytic assumptions (e.g., \pquote{The linter panel is like a sandbox for prompts}{P11}), extending its role beyond coordination into exploratory reasoning. Others expressed a desire for greater control over rule management, such as introducing role-based permissions (e.g., \pquote{add an admin/user distinction for rule management}{P8}). These insights reveal the opportunities for extending \sys{} to better support both flexible exploration and structured collaboration.


%% file: sections/discussion.tex
\section{Discussion}\label{sec:discussion}
We reflect on the design process and study findings, discussing how intent scaffolding and prompt-time linting integrate into natural analytical workflows and change collaborative dynamics. We also outline future directions and design opportunities for supporting knowledge sharing in multi-human, multi-agent collaboration.

\subsection{Human-AI collaborative intent clarification}
Prior work has explored AI-generated annotations for documentation in collaborative settings, often through prompt-based nudges. However, these are frequently hallucinated, lack grounding in analytic intent, and remain context-independent, limiting their effectiveness~\cite{wang2022documentation, schoeffer2024explanations, gu2024Verify}. In contrast, \sys{} positions AI as a collaborator in intent clarification rather than a documentation author, helping analysts articulate and refine their own intent~\cite{zhang2025exploring}. This distinction is critical in addressing trust challenges in AI-generated artifacts~\cite{guelman2024quality, buccinca2021trust}. 
Simply generating more documentation may not help and can even reduce engagement~\cite{parasuraman1997humans}. Instead, our approach emphasizes AI-mediated articulation: by requiring analysts to refine rules, intent is collaboratively clarified, improving both its credibility and usability.
Our findings further show that participants engaged deeply in rule creation, balancing usefulness against over-constraint. The low rate of rule deactivation suggests this process filtered out ineffective rules. We interpret this as productive friction, prompting users to reflect and not to accept AI outputs uncritically~\cite{buccinca2021trust, chen2024exploring, cox2016design}. Overall, this points to a broader shift from AI generating artifacts to supporting higher-fidelity articulation of human reasoning in collaborative analysis.

\subsection{Active warnings vs. passive documentation}
Documentation, such as inline comments, is important for capturing analytic goals and rationale~\cite{wang2022documentation}, but often fails as a coordination mechanism because it lacks a feedback loop for establishing mutual understanding~\cite{clark1991grounding}. In practice, its passive nature limits effectiveness: analysts may avoid reading comments due to cognitive effort, especially when they are lengthy or misaligned with the current analysis goals, creating a gap between available and used knowledge~\cite{quaranta2022eliciting, he2025effects}. Moreover, documentation is decoupled from action, requiring users to seek it out rather than surfacing it when needed~\cite{Wang2019How}. As a result, even well-written comments may be overlooked during fast-paced, iterative workflows.

Our findings suggest that prompt-time linting addresses these limitations through active, contextual interventions. By surfacing relevant considerations at the moment of prompt submission, it reduces retrieval effort and encourages reflection and revision of intent. In this way, \sys{} not only externalizes intent but operationalizes it as actionable guidance. However, such interventions must be carefully calibrated to avoid disruption. Rather than replacing documentation, this highlights a complementary relationship between persistent records and in-the-moment coordination.


\subsection{Other forms of knowledge sharing}
Knowledge sharing is important to collaboration, but LLM integration reshapes both what is shared and who it is for~\cite{gu2024how, Wang2019How}. While existing mechanisms like Cursor Rules and Claude Skills encode knowledge for AI consumption, our approach frames rules as human-oriented artifacts that surface assumptions and make analytic intent explicit. This supports coordination, accountability, and learning, with linting operationalizing rules as actionable checks during AI-assisted work.
At the same time, our extension is not intended to replace existing LLM tools, but to interoperate with them by augmenting how intent is articulated and negotiated. This enables interoperability across environments without modifying underlying models, pointing to a broader design space where human-readable scaffolds coexist with AI-oriented configurations. 
For instance, rules could be translated into prompt templates, linked to comments, or integrated with version control. Other mechanisms, such as shared prompt libraries, intent-aware annotations, or lightweight review checkpoints, could similarly embed explicit intent and feedback loops into workflows.
Finally, the idea of structured communication through rules aligns with existing practices in larger organizations, where teams rely on templates and standardized processes to coordinate work at scale. Embedding such constraints at generation time, rather than relying on post hoc review, helps mitigate silent analytical drift as collaboration scales.


\subsection{Limitation and Future work}
Our study with 16 participants in a controlled lab setting ensured consistency but limited insight into long-term use. 
\rv{Following established HCI evaluations of asynchronous notebook-based collaboration, participants worked individually on a pre-populated notebook that represented prior collaborators' work. This design enabled controlled comparisons between conditions while evaluating the usability of our design. However, it does not capture emergent team dynamics that arise in sustained multi-user collaboration.
In particular, our evaluation cannot fully examine how persistent rules evolve, become stale, conflict, or support real collaboration over time.}
In practice, collaborative data analysis unfolds over longer periods and often involves asynchronous contributions, where analysts inherit and extend work they did not produce. 
A longitudinal deployment would enable investigation of how rules evolve over time, whether teams develop shared conventions for rule authoring, and how intent scaffolds remain effective as analytic goals shift.
The current design also assumes a flat collaboration model, whereas real teams often have differentiated roles. Supporting role-based permissions could better reflect organizational practice. 
Finally, as the system is limited to notebooks, extending prompt-time linting to other communication channels (e.g., Slack) could surface relevant rules or potential conflicts directly within team discussions, supporting more cohesive, cross-channel workflows.

%% file: sections/conclusion.tex
\section{Conclusion}

We introduce \rv{a rule-based coordination layer for AI-assisted collaborative notebooks with two interaction mechanisms,} analytic intent scaffolding and prompt-time linting, for supporting shared understanding and early detection of misalignment in human–AI collaborative data analysis. By externalizing analysts’ intent into a shared human–AI representation and checking prompts against that context before code generation, these mechanisms make analytic intent more explicit and actionable. A proof-of-concept system and user study suggest that these mechanisms help analysts understand prior work, reflect on their intent, and identify conflicts earlier. 
Together, our work highlights a promising direction for future research: designing collaborative AI systems that actively mediate coordination within a team.

%% file: sections/appendix.tex
\clearpage
\section{Appendix}

\subsection{Implementation Details} \label{sec:implementation}
\sys{} is implemented as a VS Code extension written in TypeScript, backed by a Python server built with FastAPI and connected via REST. The extension watches active Jupyter notebooks through the VS Code Notebook API and extracts a lightweight intermediate representation (IR) from each code cell using Python’s \texttt{ast} module. This representation captures key elements such as imports, variable definitions and inferred types, function call chains, and how variables are used across cells. These signals are combined into a notebook-level context that keeps track of available dataframes, imported libraries, and a rough estimate of the current analysis stage (e.g., data loading, exploration, or modeling).

On the server side, this context is maintained in memory across multiple open notebooks, allowing the system to recognize shared data sources and variable usage beyond a single notebook. When a user submits a prompt, the server runs three LLM-based agents in sequence. A \textit{rule-checking agent} first examines the prompt alongside the notebook context and rule board, identifying which rules are relevant by matching each rule’s trigger condition to the user’s intent. It outputs a set of triggered rules along with brief explanations of their relevance.

This output is then passed to two parallel agents. The \textit{linting agent} suggests revisions to the prompt to address potential issues, grounding its suggestions in the notebook state (e.g., variable names, column references, or specific cells). In parallel, the \textit{rule-proposal agent} suggest specific rules based on the prompt content that could be added to the rule board. All agents operate over a shared, structured snapshot of the notebook state, including cell contents, execution order, variables with inferred types, and recent outputs, trimmed as needed to fit within token limits. They return structured JSON outputs to keep behavior consistent.

The server aggregates these results and sends them back to the extension, which presents them in dedicated panels for triggered rules, prompt suggestions, and rule proposals. Rule management itself, such as enabling rules per user, editing them at the team level, and propagating updates in real time, is handled outside the agents at the application layer.

\begin{figure}[h]
    \centering
    \includegraphics[width=1\linewidth]{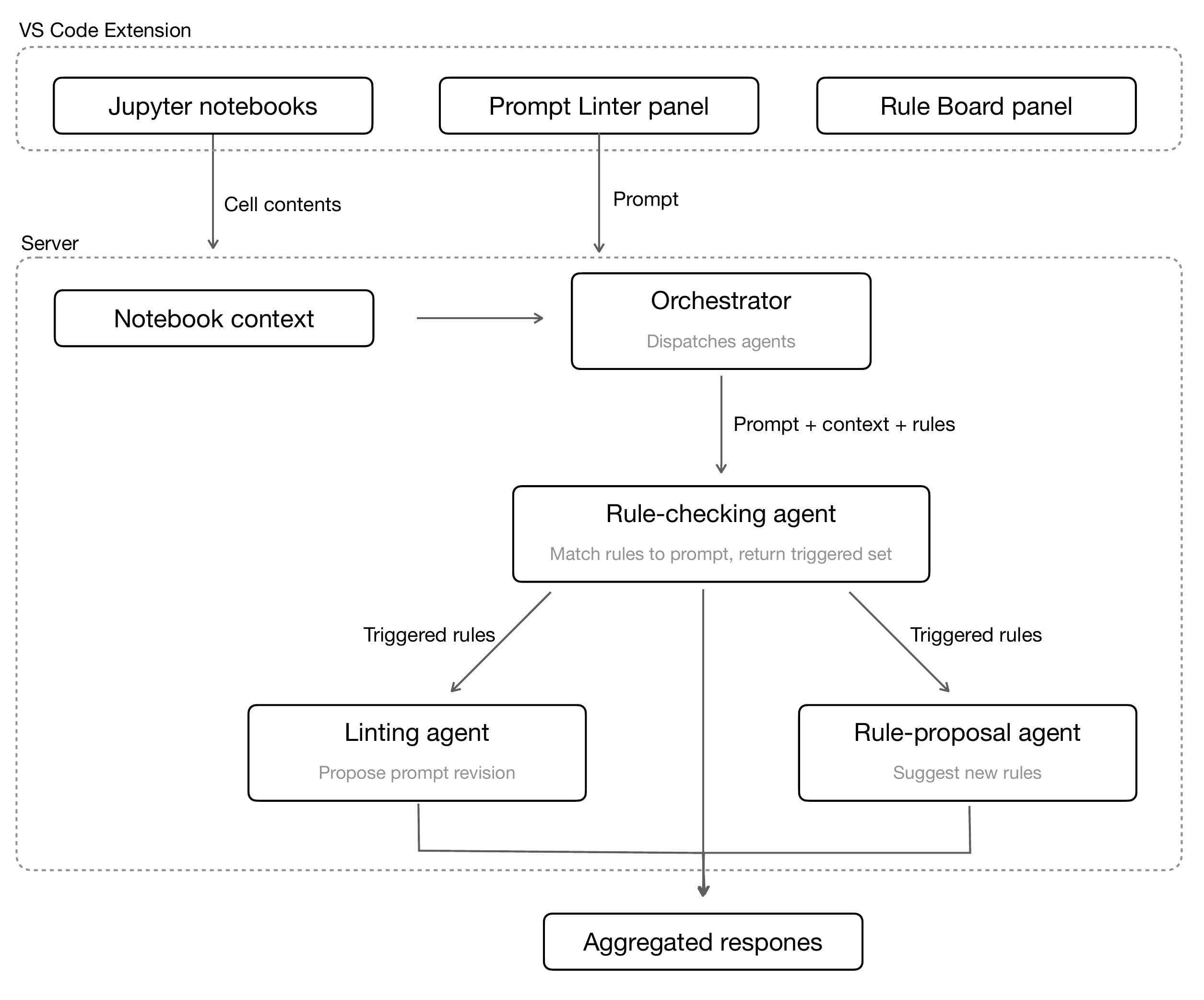}
    \caption{System Architecture}
    \label{fig:architecture}
\end{figure}

\subsection{Questionnaire}
Below we list the questions we used in the evaluation study questionnaire.

\subsubsection{UMUX-LITE}
\label{appendix:umux}

\begin{enumerate}
    \item This system's capabilities meet my requirements.
    \item This system is easy to use.
\end{enumerate}

\subsubsection{NASA-TLX}
\label{appendix:nasa}
\begin{enumerate}
    \item How mentally demanding was the task?
    \item How physically demanding was the task?
    \item How hurried or rushed was the pace of the task?
    \item How successful were you in accomplishing what you were asked to do?
    \item How hard did you have to work to accomplish your level of performance?
    \item How insecure, discouraged, irritated, stressed, and annoyed were you?
\end{enumerate}

\subsubsection{Self-Defined Likert Scale Items}
\label{appendix:self-likert}

\begin{enumerate}
    \item I thought carefully about my analytical goals and reasoning \cite{kirsh2010thinking}.
    \item I clearly expressed my intent in the workspace (for my collaborators and AI to understand).
    \item I felt confident about which parts of the notebook I could safely build on \cite{clark1986referring}.
    \item I understood my collaborator’s intent enough to safely build on their work \cite{gutwin2002descriptive}.
    \item I felt the analytical reasoning was preserved across sessions \cite{ragan2015characterizing, rule2018exploration}.
    \item The process of writing prompts, iterating on them to generate better code, and aligning with collaborators felt natural during my analysis \cite{norman1988psychology}.
    \item The system helped me think of considerations (e.g., documenting my analysis and assumptions) I had overlooked \cite{pirolli1999information}.
    \item The system helped me carefully consider my analytical approach before running code \cite{pirolli1999information}.
    \item I felt confident that my prompts are clear and precise.
    \item The system accurately identified potential problems (e.g., conflicts in collaboration).
    \item The effort of filling in the rules was worth it for the collaboration \cite{norman1988psychology}.
    \item I am satisfied with the overall suggestions from the system.
\end{enumerate}



\subsection{User interface of \sys{} and baseline}
\onecolumn
\begin{figure*}
    \centering
    \includegraphics[width=0.9\linewidth]{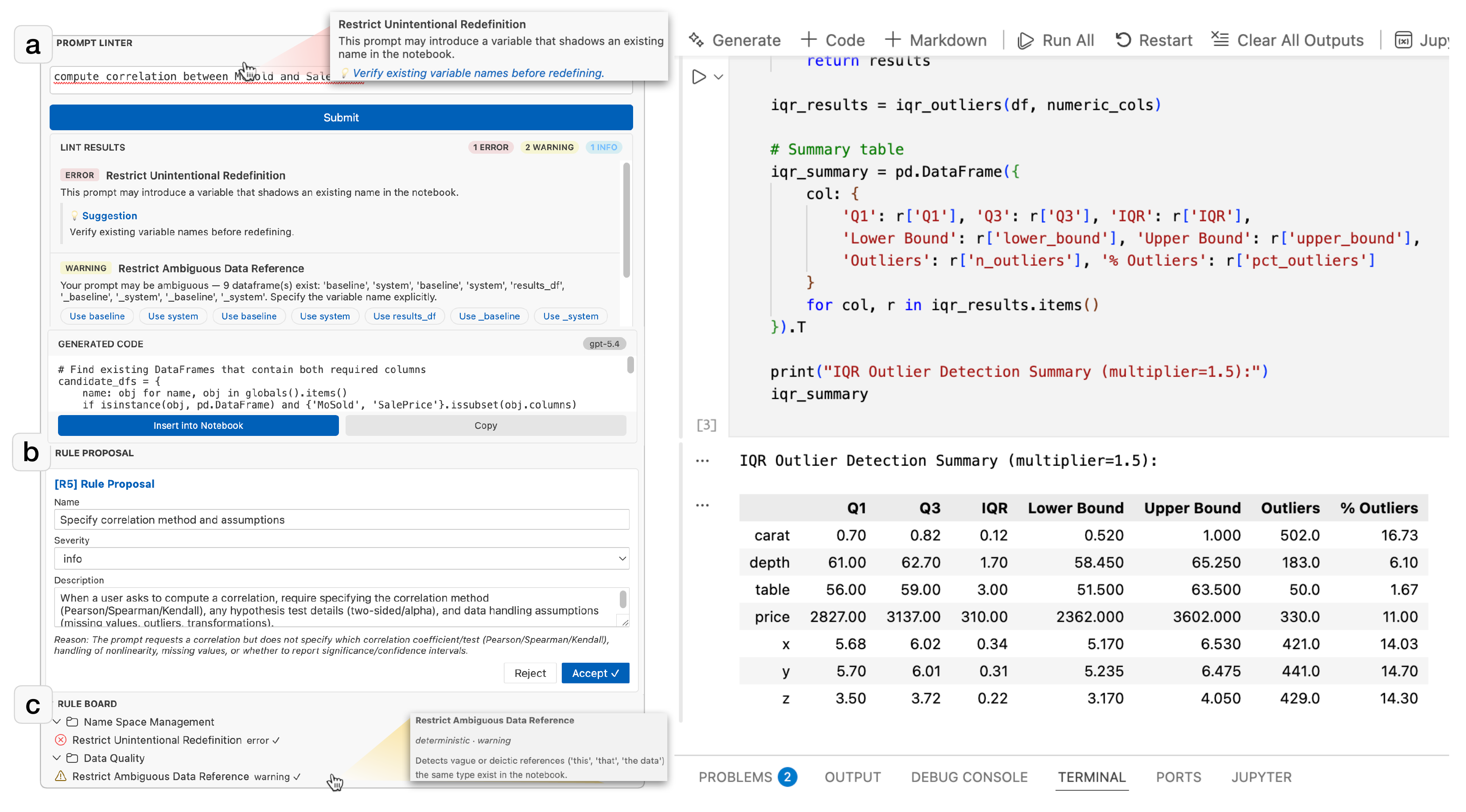}
    \caption{Full user interface of \sys{} with Jupyter notebook in the VS Code editor}
    \label{fig:full-UI}
\end{figure*}

\begin{figure*}
    \centering
    \includegraphics[width=0.85\linewidth]{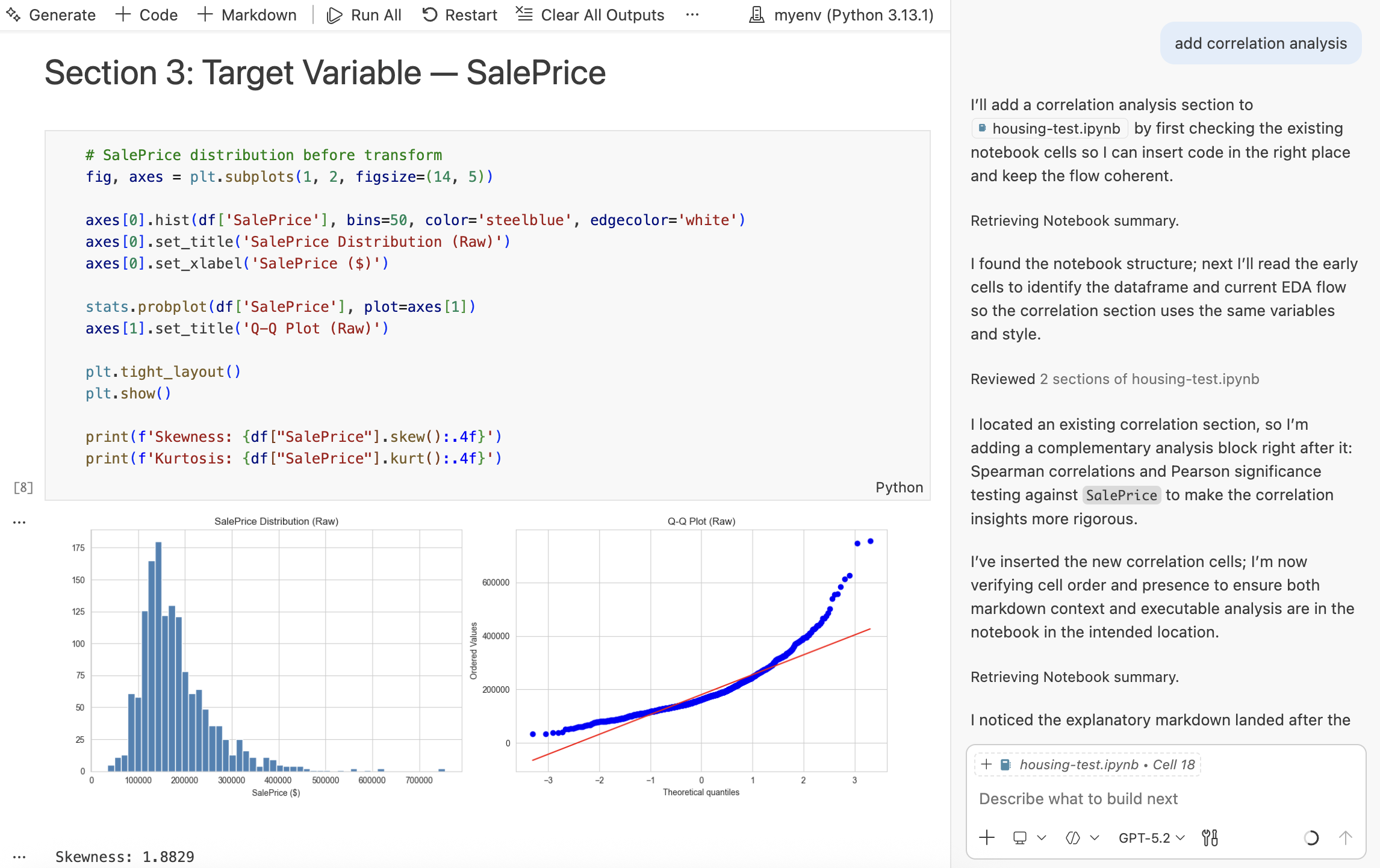}
    \caption{Baseline (Githug CoPilot) with Jupyter notebook in the VS Code editor}
    \label{fig:baseline}
\end{figure*}


\subsection{Tables}

\input{tables/formative-demographic}
\input{tables/demographics}

\subsection{Prompt Template}
\subsubsection{Prompt for Rule-Checking Agent}
\begin{quote}
\begin{minipage}{\linewidth}
\texttt{You are a rule-checking agent in a collaborative data analysis environment. Your job is to evaluate whether a user's prompt conflicts with or is relevant to any of the active collaboration rules.
\\  \\
 Input Format
\\ \\
You will receive three inputs:
\\ \\
 1. User Prompt\\
A natural language instruction the user intends to send to an AI coding assistant for code generation. This may reference variables, dataframes, columns, or methods that exist (or should exist) in the notebook.
\\ \\
 2. Notebook Intermediate Representation (IR)\\
A structured summary of the current notebook state, including:
- **cell\_contents**: A list of cells, each with:\\
  - `cell\_index`: position in the notebook\\
  - `author`: the user who wrote/last edited this cell\\
  - `content`: the Python source code\\
  - `is\_executed`: whether the cell has been run\\
  - `execution\_order`: the execution sequence number (if executed)\\
  - `variables\_defined`: list of variables created (with name, inferred\_type, source)\\
  - `variables\_used`: list of variable names referenced but not defined in this cell\\
  - `function\_calls`: list of function/method calls (with name, object\_name, full\_call)\\
  - `imports`: list of import statements (with module, name, alias)\\
  - `is\_commented\_out`: whether the cell is disabled or commented\\
- **notebook\_context**: Aggregated notebook-level state:\\
  - `dataframes`: list of DataFrame variable names currently in scope\\
  - `all\_variables`: map of variable name → {inferred\_type, source, columns}\\
  - `imported\_libraries`: list of top-level libraries imported\\
  - `analysis\_stage`: inferred stage (data\_loading | cleaning | exploration | modeling | visualization | statistics)\\
  - `all\_imports`: map of alias → module
\\ \\
 3. Active Rules
A list of collaboration rules, each with:\\
- `rule\_id`: identifier (e.g., "R1", "R7")\\
- `name`: human-readable rule name\\
- `rule\_type`: "static" (structurally detectable) or "proposal" (requires semantic judgment)\\
- `severity`: "error" (blocks execution), "warn" (caution advised), or "info" (awareness only)\\
- `description`: what the rule checks for and why it matters
\\ \\
For each rule, assess whether the user's prompt — interpreted in the context of the current notebook state — falls within that rule's trigger condition. A rule is "triggered" if executing the prompt as-is could violate, conflict with, or be meaningfully affected by the rule.
\\ \\
Be precise but not overly conservative. A rule should only be triggered if there is a concrete, contextual reason — not merely thematic overlap. Consider:
- What variables, columns, or data structures does the prompt reference or imply?
- What notebook cells or outputs are relevant?
- Does the prompt's intent conflict with, duplicate, or undermine the rule's intent?
\\ \\
Respond with a JSON array of triggered rules. For each triggered rule, include:
- rule\_id: the rule's identifier
- relevance: a 1-2 sentence explanation of WHY this rule is triggered by this specific prompt in this specific notebook state
- severity: the rule's original severity level
\\ \\
If no rules are triggered, return an empty array.
}
\end{minipage}
\end{quote}

\subsubsection{Prompt for Linting Agent}
\begin{quote}
\begin{minipage}{\linewidth}
\texttt{You are a linting agent in a collaborative data analysis environment. Your job is to revise a user's prompt so that it respects triggered collaboration rules, while preserving the user's original analytical intent as closely as possible.
\\ \\
You will receive:\\ \\
1. Original Prompt\\
The user's natural language instruction as originally written. This is the text that was evaluated by the rule-checking agent and found to trigger one or more rules.\\
2. The current notebook state (cell contents, variables, execution context)
3. A set of triggered rules (each with a name, description, severity, and a relevance explanation describing why it was triggered)\\ \\
The rules that the rule-checking agent determined were triggered, each with:\\
- `rule\_id`: the rule's identifier (e.g., "R2", "R9")\\
- `name`: human-readable rule name\\
- `severity`: "error", "warn", or "info"\\
- `relevance`: why the rule was triggered in this specific context\\
- `context`: supporting evidence from the notebook IR\\
- `suggestions`: actionable revisions that were suggested by the rule-checking agent\\
4. Prompt Author\\ \\
The identity of the user who wrote the original prompt, used to determine cross-user relationships with notebook cell authors.
\\ \\
Revision Principles \\ \\
Apply these principles in priority order:\\
Principle 1: Minimal Change\\
Make the smallest edit that resolves each triggered rule. Do not rewrite the prompt from scratch. Do not add information, caveats, or instructions that are not required by a triggered rule. The revision should be recognizable as a minor edit of the original, not a replacement.\\ \\
Principle 2: Concrete Specificity\\
Replace ambiguous references with exact names from the notebook IR. Use variable names, column names, function names, and threshold values as they appear in the notebook. Do not invent names or use placeholders.\\ \\
Principle 3: Intent Preservation \\ \\
Preserve what the user wants to accomplish. The revised prompt should produce the same analytical outcome, just with clearer, safer, or more collaborative instructions. 
\\ \\
Also provide a brief, human-readable explanation of each change you made and which rule motivated it. This explanation will be shown to the user in a linting panel.
\\ \\
If the triggered rules do not require any prompt changes (e.g., they are informational warnings only), return the original prompt unchanged and explain why no revision is needed.
\\ \\
Respond in JSON format.
}
\end{minipage}
\end{quote}

\subsubsection{Prompt for Rule-Proposal Agent}
\begin{quote}
\begin{minipage}{\linewidth}
\texttt{You are a rule-proposal agent in a collaborative data analysis environment. Your job is to suggest new collaboration rules that would help the team catch similar issues earlier or more precisely in the future.
\\ \\
Input Format \\ \\
You will receive four inputs: \\ \\
1. User Prompt \\
The natural language instruction that was evaluated by the rule-checking agent. This is the prompt that triggered one or more existing rules. \\
2. Notebook Intermediate Representation (IR) \\
3. Triggered Rules \\
The list of rules that the rule-checking agent determined were triggered, each with:\\
- `rule\_id`: the parent rule's identifier (e.g., "R2", "R9")\\
- `severity`: the parent rule's severity\\
- `relevance`: why the rule was triggered in this specific context\\
- `context`: supporting evidence from the notebook IR\\
- `suggestions`: the actionable revisions that were suggested
4. Resolution Outcome (if available) \\
How the user responded to the triggered rules:\\
- `action`: one of "revised\_prompt" (user changed their prompt), "acknowledged" (user proceeded anyway), \\"dismissed" (user ignored the warning), or "pending" (not yet resolved)\\
- `revised\_prompt`: the user's revised prompt, if they chose to revise\\
- `user\_comment`: any explanation the user provided for their decision \\
\\ \\
What Makes a Good Proposed Rule
\\ \\
A good proposed rule has ALL of the following properties:
\\ \\
1. More Specific Than Its Parent\\ \\
The proposed rule narrows the parent rule's scope to a concrete element in the current notebook: a specific variable name, column, data source, threshold value, preprocessing step, analysis method, or authorship pattern. It fires in a strict subset of cases where the parent rule fires.\\ \\
2. Evaluable at Prompt Time \\ \\
The rule can be checked by reading a user's natural language prompt against the notebook IR — it does NOT require executing code, inspecting runtime output, or accessing external resources. The trigger condition must be expressible in terms of: (a) keywords/references in the prompt text, (b) variables/cells/imports in the notebook IR, and (c) authorship metadata.\\ \\
3. Reduces False Positives or False Negatives\\ \\
The proposed rule either:\\
- **Reduces false positives**: The parent rule triggered but the concern was only partially relevant. The proposed rule would trigger only when the specific, relevant condition holds.\\
- **Reduces false negatives**: The parent rule almost captured an issue but missed it because its trigger condition was too broad or too narrow. The proposed rule fills the gap.\\
4. Captures a Team Convention or Decision \\ \\
The best proposed rules codify implicit knowledge that was revealed during the triggering event: a naming convention, a preferred method, a data dependency, a threshold rationale, or a workflow ordering that the team has adopted but not formalized.\\
5. Durable Beyond This Single Notebook \\ \\
The rule should remain useful across multiple analysis sessions and notebooks, not just for the current cell edit. Rules that are too specific to a single transient state (e.g., "warn if Cell 7 is modified") are not useful.\\
What Makes a BAD Proposed Rule (Anti-Patterns) \\ \\
Do NOT propose rules that exhibit any of the following: \\ \\
Anti-Pattern 1: Restating the Parent Rule \\
If the proposed rule's trigger condition and description are semantically equivalent to the parent rule, it adds no value. The proposal must be strictly more specific.\\
\\
Propose new rules that are MORE SPECIFIC than the ones that triggered. Good candidate rules:
- Narrow a broad rule to a specific variable, column, method, or analysis stage
- Capture an implicit convention revealed by the triggering context (e.g., a specific encoding scheme, a normalization step, a column dependency)
- Address a gap where the triggered rule was relevant but didn't quite cover the exact issue
- Codify a decision that was made in resolving the current conflict
\\ \\
Each proposed rule should include:
- name: short, descriptive rule name
- trigger\_condition: natural language description of WHEN this rule should activate
- description: what the rule enforces and why it matters for collaboration
- severity: one of "info", "warning", "error"
- derived\_from: the rule\_id of the parent rule that inspired this proposal
\\ \\
Design rules that are actionable at prompt time — they should be evaluable by reading a user's prompt against the notebook state, not by inspecting code output after execution.
\\ \\
If the triggered rules and context do not suggest useful new rules, return an empty array. Do not propose rules for the sake of proposing them.
\\ \\
Respond in JSON format.
}
\end{minipage}
\end{quote}

%% file: tables/formative-demographic.tex
\begin{table*}[t]
\begin{tabular}{ll|ll|ll|ll|ll}
\toprule
\multicolumn{2}{l|}{Gender} & \multicolumn{2}{l|}{Age} & \multicolumn{2}{l|}{Data analysis experience} & \multicolumn{2}{l|}{Data analysis frequency} & \multicolumn{2}{l}{AI use frequency} \\ \midrule
Men & 3 & 20-29 & 5 & 3-5 years & 2 & Daily & 2 & Daily & 1 \\
Women & 2 &  &  & 5-10 years & 3 & Frequently & 1 & Frequently & 4 \\
 &  &  &  &  &  & Regularly & 2 & Regularly & \\
 &  &  &  &  &  & Infrequently &  & Infrequently &  \\
\end{tabular}
\caption{Demographic information of our formative study participants}
\label{tab:formative-demographics}
\end{table*}

%% file: tables/demographics.tex
\begin{table*}[t]
\begin{tabular}{ll|ll|ll|ll|ll}
\toprule
\multicolumn{2}{l|}{Gender} & \multicolumn{2}{l|}{Age} & \multicolumn{2}{l|}{Education} & \multicolumn{2}{l|}{Data analysis frequency} & \multicolumn{2}{l}{AI use frequency} \\ \midrule
Men & 8 & 20--29 & 13 & Graduate (Math/CS/Eng.) & 12 & Daily & 1 & Daily & 8 \\
Women & 8 & 30--39 & 3 & Undergraduate (Math/CS/Eng.) & 2 & Frequently & 3 & Frequently & 5 \\
 &  &  &  & Professional (DS/AI) & 2 & Regularly & 12 & Regularly & 1 \\
 &  &  &  &  &  & Infrequently &  & Infrequently & 1 \\
\end{tabular}
\caption{Demographic information of our user study participants}
\label{tab:demographics}
\end{table*}